\documentclass[prl,aps,superscriptaddress,noshowpacs,twocolumn,reprint,noshowkeys,amssymb,amsmath,amsfonts]{revtex4-1}

\usepackage{graphicx}
\usepackage{theorem}
\usepackage{dcolumn}
\usepackage{longtable}
\usepackage{bm}
\usepackage{accents}
\usepackage{natbib}

\newcommand{\Ee}{{\rm e}}
\newcommand{\Dd}{{\rm d}}

\newcommand{\Ii}{{\rm i}}

\newcommand{\ie}{{\textit{i.e.}}}

\newcommand{\etc}{{\textit{etc.}}}
\usepackage{xcolor,xspace}

\makeatother

\makeatletter
\renewcommand\@make@capt@title[2]{%
        \@ifx@empty\float@link{\@firstofone}{\expandafter\href\expandafter{\float@link}}%
        \sffamily{\textbf{#1}}\@caption@fignum@sep#2
}%

\makeatother

\begin{document}

\title{Quantum optics of single electrons in quantum liquid and solid helium-4}

\author{Matthew Otten}
\author{Xianjing Zhou}
\author{Xufeng Zhang}
\author{Dafei Jin}
\email{djin@anl.gov}
\affiliation{Center for Nanoscale Materials, Argonne National Laboratory, Argonne, Illinois 60439, USA}

\date{\today}

\begin{abstract}

Single electrons can be conceived as the simplest quantum nodes in a quantum
network. Between electrons, single photons can act as quantum channels to
exchange quantum information. Despite this appealing picture, in conventional
materials, it is extremely difficult to make individual electrons and photons
coherently interact with each other at the visible-infrared wavelengths
suitable for long-distance communication. Here we theoretically demonstrate that
the self-confined single-electron structure in condensed helium-4 can be a
fascinating candidate for single-electron quantum nodes. Each electron in helium
forms a bubble of 1 to 2~nm radius and coherently interacts with mid-infrared
photons. A parametrically amplified femtosecond laser can drive the electrons
into any superposition between the ground and excited states. An electron
inside a slot-waveguide cavity can strongly couple with cavity photons and
exhibits vacuum Rabi oscillations. Two electrons in the cavity naturally
generate entanglement through their respective coupling to the lossy cavity. The
electron-in-helium system offers unique insight in understanding nonequilibrium
quantum dynamics.

\end{abstract}

\maketitle
\pretolerance=8000 

A quantum network consists of quantum nodes and quantum channels~\cite{kimble2008quantum}. Quantum
channels achieve remote quantum information exchange through the universal
``flying" qubits~---~photons~\cite{tanzilli2005photonic,o2009photonic}, whereas quantum nodes perform local quantum
information processing via various ``trapped" qubits~---~ions, atoms, molecules,
defect centers, quantum dots, \etc~\cite{weber2010quantum,de2012quantum,thiele2014electrically,muhonen2014storing,saffman2016quantum,brown2016co}. Beyond these, perhaps the most exotic but simplest quantum node conceivable is a single electron~\cite{koppens2006driven,mi2017strong}.
Unfortunately, single electrons in free space or semiconductor junctions
do not resonantly interact with visible-infrared (Vis-IR) photons, and cannot
be directly linked to long-distance quantum communication~\cite{tanzilli2005photonic,o2009photonic}.

Here we propose a single-electron
quantum-node platform, based on an extraordinary electronic structure
in condensed, quantum liquid and solid helium-4 (He-4). An excess electron
injected into He-4 spontaneously forms a 1-2~nm radius self-confined
structure, called an electron bubble~\cite{sommer1964liquid,cohen1969electron,moroshkin2008atomic,sabouret2008signal}, which resembles the textbook example of a quantum particle confined in a finite-depth spherical well. The electron can resonantly
interact with mid-infrared (mid-IR) photons and the wavelength can be tuned by
changing the ambient pressure~\cite{grimes1992infrared,golov1994ir,maris2003properties,jin2010electrons}. We find that a femtosecond laser extended into the mid-IR
regime by an optical
parametric amplifier (OPA) can drive the electron into Rabi
oscillations and prepare its wavefunction into arbitrary superpositions between the ground and
excited states. A single electron situated inside an on-chip slot-waveguide cavity can
strongly couple with cavity photons and exhibit vacuum Rabi oscillations with microseconds lifetime. Two electrons placed in the cavity
automatically generate
entanglement with each other through their coupling to the lossy cavity.
Depending on cavity's photon loss rate, the entangled states show varied
population and concurrence. The electron-in-helium system opens new
opportunities for studying coherent nonequilibrium
quantum dynamics of a quantum particle embedded in quantum
matter~\cite{leggett1987dynamics}.

\noindent\textbf{Condensed helium-4 and electron bubble}

\begin{figure*}[tbh]
\centerline{\includegraphics[scale=0.9]{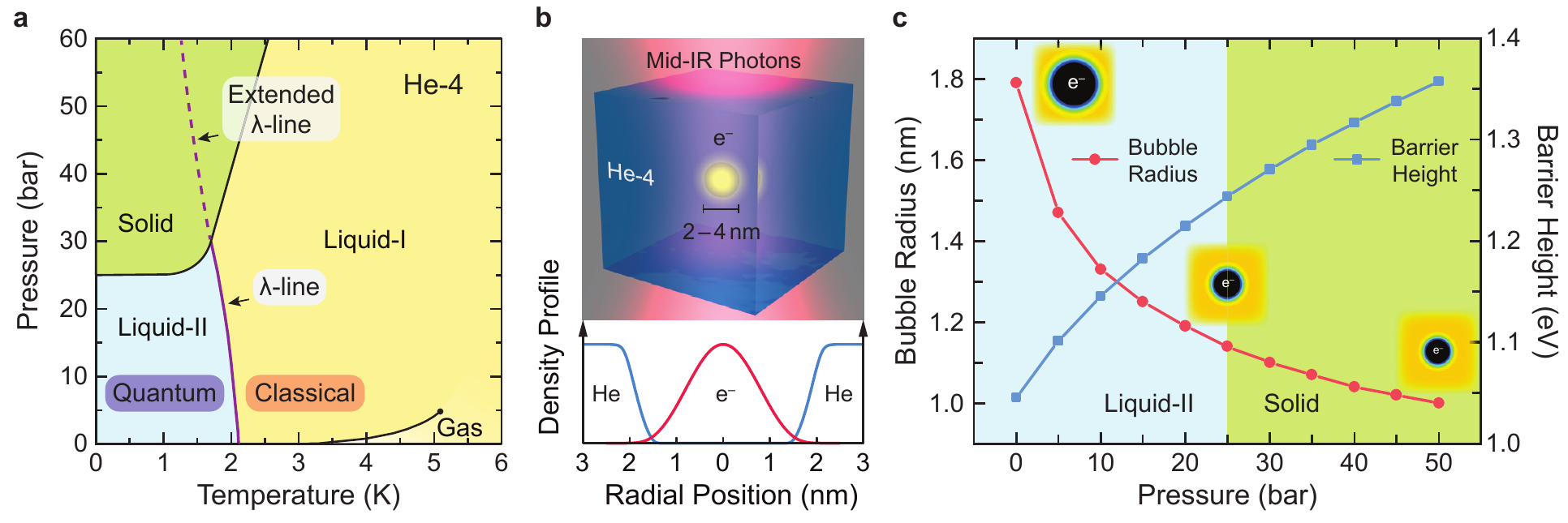}}
\caption{Self-confinement of a single electron in condensed He-4. \textbf{a.} Phase
  diagram of He-4. The $\lambda$-line (and its extension) separates the quantum
  and classical matter regimes in both the liquid and solid phases. \textbf{b.}
  Upper panel: Schematics of a spherical ground-state electron bubble under optical exposure of mid-infrared (mid-IR) photons. The diameter of the bubble is tunable between 2 to 4~nm under 0 to 50~bar pressure. Lower pannel: Density-functional calculated helium and electron density profile at zero pressure. \textbf{c.} Ground-state bubble radius $R$ and potential barrier height $U$ from the bulk He-4 versus pressure at near-zero temperature. The insets display the color plots of calculated helium density around the bubble at 0, 25, and 50~bar pressure. }\label{Fig:HeliumAndElectronStructure}
\end{figure*}

Liquid and solid He-4 are the purest condensed matter in nature. Crossing the
famous $\lambda$-transition at about 2~K temperature (depending on the
pressure), condensed He-4 changes from a classical matter to a quantum matter~\cite{wilks1967properties,leggett2006quantum}, as shown in Fig.~\ref{Fig:HeliumAndElectronStructure}a. In the quantum
regime, He-4 takes a quantum liquid-II (superfluid) phase below about 25~bar and
a quantum solid phase above~\cite{werner2004liquid,wilks1967properties}.

When an excess electron is injected into liquid or solid He-4, it repels the surrounding helium atoms away under the Pauli exclusion and forms a three-dimensional (3D) nanometric
quantum well that confines its own wavefunction inside. This
structure,  as illustrated in Fig.~\ref{Fig:HeliumAndElectronStructure}b, is called an electron bubble. The
characteristic radius of the electron bubble is about 2~nm
at zero pressure in the liquid phase. It can be continuously squeezed down to
about 1~nm at 50~bar pressure in the solid phase. Since the bubble size is
much larger than He-4 atomic size ($\sim0.31$~{\AA}), the ground-state
electron bubble, either in liquid or
solid He-4, is almost perfectly spherical~\cite{moroshkin2008atomic}. The
crystal order of solid He-4 does not affect the electronic and optical
properties of electron bubble, except for the size and mobility~\cite{grimes1992infrared,golov1994ir,dionne1972effect,shikin1977mobility}.
In our present study, we include both liquid and solid He-4, focusing on the
near-zero temperature $T<1$~K and liquid-solid transition pressure $p\approx
25$~bar for most calculations.

\begin{figure}[tbh]
\centerline{\includegraphics[scale=0.9]{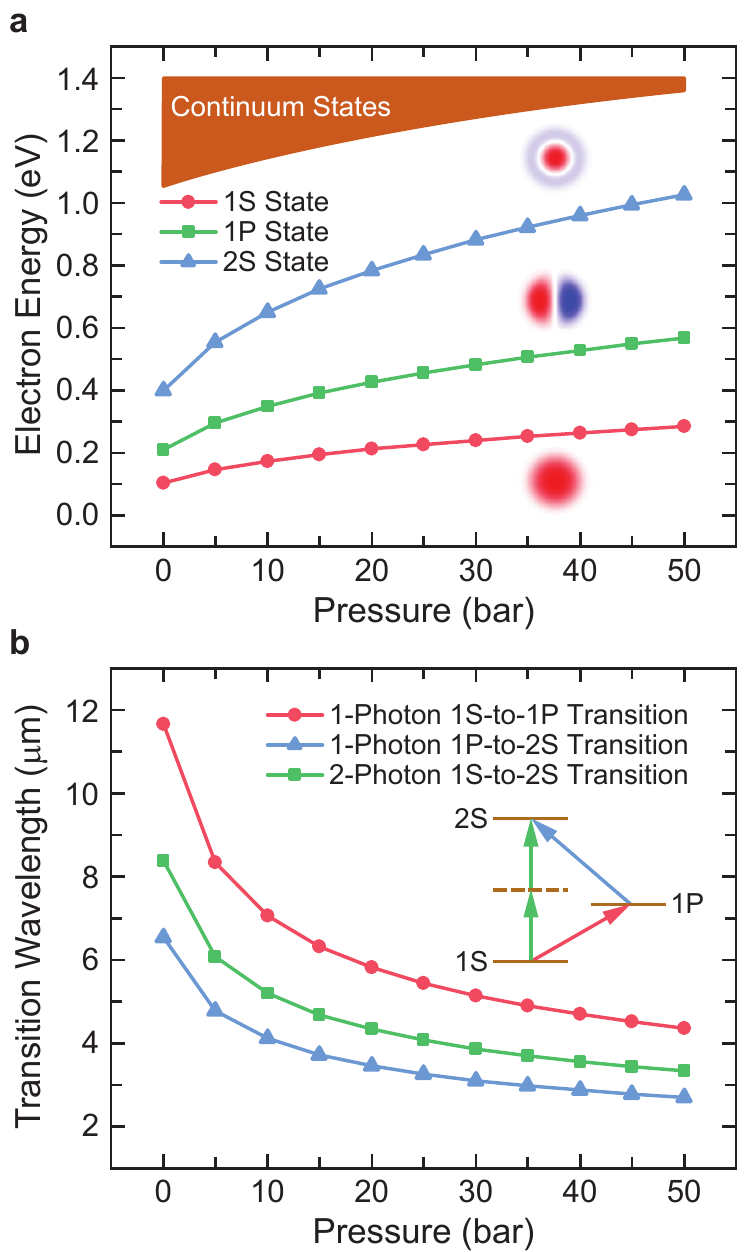}}
\caption{Electronic and optical structures of a spherical electron bubble. \textbf{a.}
  Eigenenergies and wavefunction profiles of several representative bound states: 1S, 1P and 2S, and the continuum of extended states when the electron is ejected out of the bubble. \textbf{b.} Representative optical transition wavelengths between 1S, 1P and 2S states.
  The inset illustrates the processes. }\label{Fig:BubbleEnergyAndTransition}
\end{figure}

\noindent\textbf{Electronic structure and optical transitions}

The electron confined inside the bubble is an extremely quantum object. It
bears a finite potential barrier from the condensed helium,
\begin{equation}
U(p) =  \frac{2\pi \hbar^2 l}{m_\text{e}} n_{\text{He}}(p),
\end{equation}
where $l\approx 1$~{\AA} is the s-wave scattering length between an electron and
a helium atom, $n_{\text{He}}$ is the bulk helium number density as a function of pressure, and $U$ is the barrier height, which increases from about 1 to 1.4~eV above the vacuum level as the pressure increases from 0 to 50~bar~\cite{berloff2000capture,grau2006electron,jin2010vortex}. This barrier is sufficient to generate a series of electron bound states
labeled by the angular quantum number $l=0,1,2,\dots$ corresponding to the S, P,
D, $\dots$ orbitals, and the radial quantum number $n=1,2,3,\dots$ representing
the number of nodes in the radial wavefunction~\cite{maris2003properties}. Such electronic structure is
the same as that of the textbook example of a quantum particle confined in a
finite-depth spherical well~\cite{griffiths2018introduction}. The only difference is that the potential well here is
deformable, since the electron and helium make up a self-adjusting system and the
interface results from the force balance between electron's quantum pressure and
helium's hydrodynamic pressure. The interfacial thickness between the electron
and helium is of the order of superfluid healing length $\xi\approx1$~{\AA}. In
Fig.~\ref{Fig:HeliumAndElectronStructure}c, we give the bubble radius $R$ and
barrier height $U$ from our density-functional theory (DFT) calculation (see
Methods)~\cite{jin2010electrons,jin2010vortex}, which has been attested to agree well with experiments. The
inset gives the color plot of helium density, which smoothly varies from 0 in
the bubble center to $n_{\text{He}}$ in the bulk helium. The bubble radius is
defined at the position of $\frac{1}{2}n_{\text{He}}$.

Under illumination, the electron in the bubble can transition from the 1S
ground state to many excited states. The most relevant transitions are the
1-photon 1S-to-1P, 1-photon 1P-to-2S, and 2-photon 1S-to-2S transitions. Other
transitions have much smaller probabilities
and can be safely ignored~\footnote{With the laser field strength used in our
  paper, we have calculated and confirmed that the probabilities for single- or
  multi-photon transitions onto other localized or continuum states are all
  order-of-magnitude smaller then the presented transitions.}.
Figure~\ref{Fig:BubbleEnergyAndTransition}a shows the DFT calculated
eigenenergies versus pressure for the representative
eigenstates 1S, 1P, and 2S. The typical eigenenergies are of the order of
several 100~meV. Figure~\ref{Fig:BubbleEnergyAndTransition}b shows the optical
transition wavelengths, which are in
the mid-infrared (mid-IR) regime from 3 to 12~$\mu$m. For photon energy
exceeding 1~eV, the electron can be ejected via the photoelectric effect into
the continuum states of bulk helium,
as indicated in Fig.~\ref{Fig:BubbleEnergyAndTransition}a.

\noindent\textbf{Laser-driven Rabi oscillations}

Previous studies of optical transitions of electron bubbles were limited to weak and
continuous light illumination~\cite{grimes1992infrared,golov1994ir,konstantinov2003detection}. The transition probabilities were so low
that on average only a small fraction of electrons could end in the
excited states. Today's Ti:Sapphire femotosecond
lasers equipped with an OPA and difference frequency generation (DFG) can output several microJoules of pulse energy in
mid-IR, with 100 to 200~fs pulse width at 1 to 5~kHz repetition rate~\cite{spectraphysics}. This
enables coherently driving an electron in the bubble into Rabi oscillations by
each laser pulse~\cite{scully1999quantum}.

Let us take the pressure $p=25$~bar. The calculated bubble radius is
$R=1.14$~nm, the 1-photon 1S-to-1P transition wavelength is
$\lambda=5.44$~$\mu$m, and electric dipole moment is $d=0.40$~e$\cdot$nm. The
output pulse energy from an OPA at this wavelength can be as high as 5~$\mu$J~\cite{spectraphysics}. To emulate a realistic experiment, we assume only 10\% photons can reach
the center of helium cell after multiple reflection on the cryostat windows.
This number of photons can be focused into a 100~$\mu$m diameter area by a
silicon lens. The electric field at the center can be as strong as
$\mathcal{E}\approx 0.1$~V~nm$^{-1}$, more than enough to produce a few cycles
of Rabi oscillations in sub-picoseconds.

\begin{figure*}[htb]
\centerline{\includegraphics[scale=0.9]{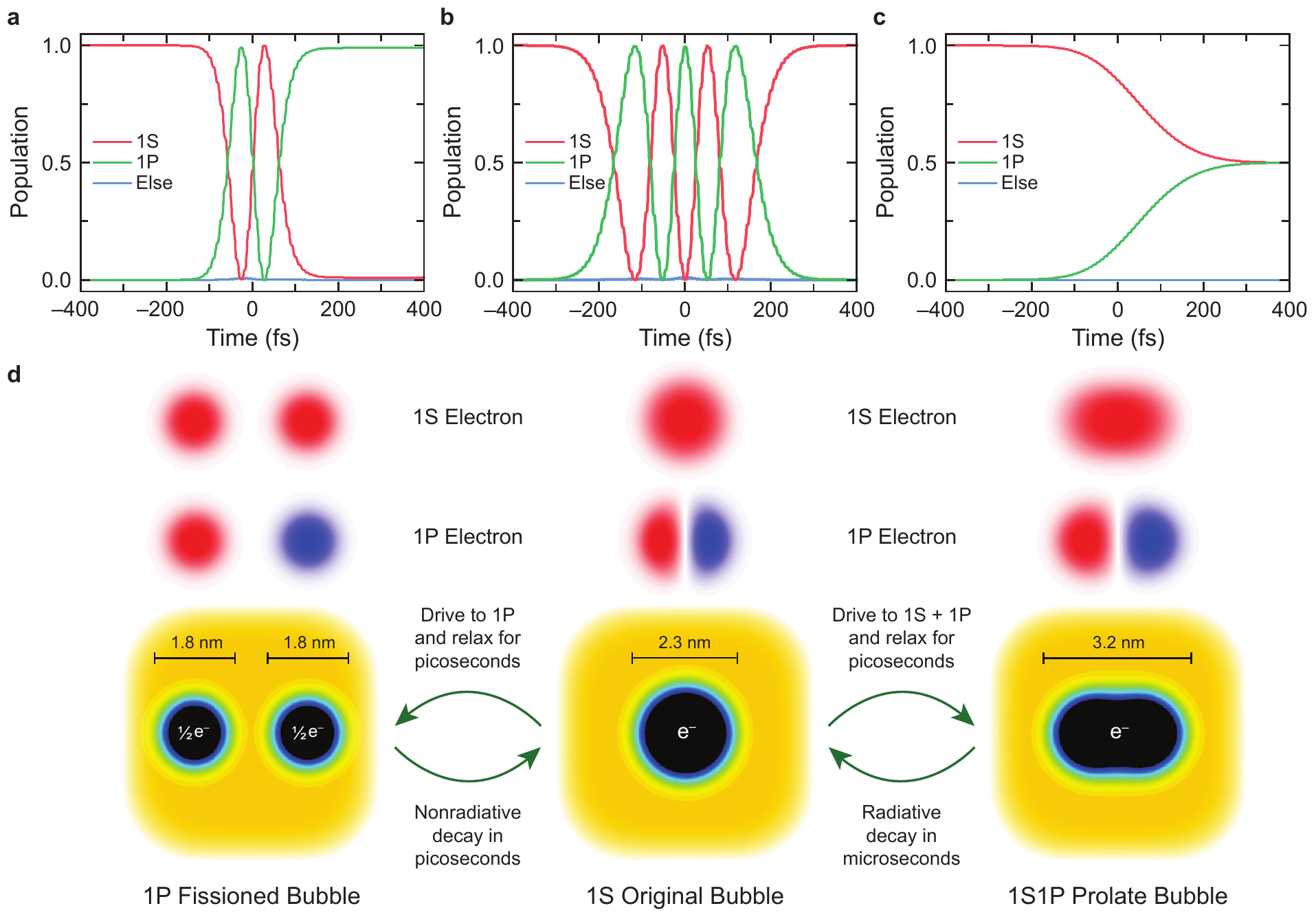}}
\caption{Laser-driven Rabi oscillations of a single electron in a spherical bubble between 1S and 1P states and the resulting bubble dynamics. \textbf{a.} Laser pulse width $\mathcal{W}=100$~fs and electric field strength $\mathcal{E}=0.1$~V~nm$^{-1}$. The end state is almost
  purely 1P. \textbf{b.} $\mathcal{W}=200$~fs and $\mathcal{E}=0.1012$~V~nm$^{-1}$. The end state
  is almost purely 1S. \textbf{c.} $\mathcal{W}=200$~fs and $\mathcal{E}=0.0085$~V~nm$^{-1}$. The end state is fine-tuned to be an equal superposition of 1S
  and 1P states. Also shown is the population leakage to all other (discrete or
  continuum) states. \textbf{d.} Versatile bubble dynamics versus different end electron
  states after the laser pulses. On the left, the bubble evolves into a
  fissioned state and goes back to the original state in picoseconds. On the
  right, the bubble evolves into a prolate shape and can stay for microseconds
  before spontaneous emission happens. In the middle, the bubble remains
  spherical.}\label{Fig:LaserDrivenRabiOscillation}
\end{figure*}

Figure~\ref{Fig:LaserDrivenRabiOscillation}a-c shows our calculated electron Rabi oscillations between the 1S and
1P states under various laser pulse configurations (see Methods). Fig.~\ref{Fig:LaserDrivenRabiOscillation}a has a pulse width
$\mathcal{W}=100$~fs and field strength $\mathcal{E}=0.1$~V~nm$^{-1}$. After one and a
quarter cycle, the electron ends up in the 1P state.
Fig.~\ref{Fig:LaserDrivenRabiOscillation}b has a slightly higher field strength
and doubled pulse width $\mathcal{W}=200$~fs. After three cycles, the
electron is brought back to the 1S state. Importantly, the pulse strength can be
tuned to fit an arbitrary superposition of 1S and 1P.
Fig.~\ref{Fig:LaserDrivenRabiOscillation}c has a pulse width
$\mathcal{W}=200$~fs and a fine-tuned field strength
$\mathcal{E}=0.0085$~V~nm$^{-1}$, leaving the electron in an equal
superposition of the 1S and 1P states.

The Rabi oscillations are much faster than the spontaneous emission in this
system. The latter takes place in microsecond time scale and is thus irrelevant
to the optical transitions during a single laser pulse. Nevertheless,
interesting phenomena happen after each laser pulse~---~the bubble shape can
deform~\cite{maris2003properties,jin2010electrons,mateo2011excited}. This is
fundamentally different from the textbook problem of a rigid
spherical well. Since the excited electron wavefunction contains nodal surfaces,
it breaks the force balance between the electron
quantum pressure and helium hydrodynamic pressure. Helium can run into the nodal
regions and redefine the electron's Hamiltonian. This process is often accompanied
with intense phonon emission~\cite{jin2010electrons,mateo2011excited}. Due to the heavy helium atom mass compared
with the electron mass ($7296:1$),
the characteristic time of shape deformation is 1-10~ps in the liquid phase
(varying with pressure) and about 1-10~ns in the solid phase~\cite{moroshkin2008atomic,jin2010electrons}. The latter is
$10^3$ slower because the mass transport in solid He-4 is by atomic hopping~\cite{dionne1972effect,shikin1977mobility}. In any case, the shape deformation is much slower than the laser pulse width. Therefore, during each pulse, the Frank-Condon principle ensures a fixed
helium profile and electronic eigenstates, but after the pulse (and before the
next pulse arrives in $\sim1$~ms), the electron bubble can undergo various
intermediate processes until it finally relaxes back to the original 1S ground
state.

Figure~\ref{Fig:LaserDrivenRabiOscillation}d illustrates several representative
intermediate bubble states corresponding to
Fig.~\ref{Fig:LaserDrivenRabiOscillation}a-c. In the middle, the electron
returns to the 1S ground state and the bubble remains spherical. No deformation
happens. On the left, the electron ends in the 1P excited state, where the quantum
pressure vanishes at the nodal surface. Helium can run into the bubble waist and
generate a double-well potential to confine two separated one-half electron
wavefunctions with antisymmetric parity. This is known as the unstable fission
state of the electron bubble~\cite{maris2000fission,maris2003properties,jin2010electrons,mateo2011excited}. This process takes several picoseconds in
liquid and a few nanoseconds in solid. However, whether this truly happens and
how the electron wavefunction loses its coherence have been open questions for
years~\cite{maris2000fission,jackiw2001fractional,flowers2001progress,konstantinov2003detection,wei2016exotic}.

An even more attractive configuration is shown on the right of
Fig.~\ref{Fig:LaserDrivenRabiOscillation}d. The electron ends in an
equal superposition of 1S and 1P states,
\begin{equation}
|\text{1S1P}\rangle = \frac{1}{\sqrt{2}}\left(\Ee^{\Ii\theta_1} \Ee^{-\Ii \omega_{\text{1S}} t} |\text{1S}\rangle + \Ee^{\Ii\theta_2} \Ee^{ -\Ii \omega_{\text{1P} } t} |\text{1P}\rangle\right),
\end{equation}
where $\theta_1$ and $\theta_2$ are initial phases. Due to the presence of a 1S
component,
the total wavefunction no longer has a stationary nodal surface. The bubble
adiabatically evolves into a prolate spheroid shape,
which is a long-lived metastable state. The new electronic eigenstates in the
bubble are the prolate 1S and 1P states. Later, we shall note that this 1S1P
state is related to the dressed states of electron and photons inside a cavity.
In the microsecond time scale, the 1P component vanishes by spontaneous
radiation and the system returns to the original 1S spherical bubble state.

Figure~\ref{Fig:LaserDrivenRabiOscillation} has only demonstrated several examples. By adjusting
the pulse energy in experiment, one can in principle bring the electron
into arbitrary superpositions of 1S and 1P states or other high-lying
states. This can lead to a variety of bubble shapes together with a variety of quantum dynamics. These
extraordinary behaviors suggest that one can use photons to tailor the
Hamiltonian and engineer electron wavefunctions in quantum matter He-4. To our
knowledge, no other systems share similar capabilities.

\begin{figure*}[tbh]
	\centerline{\includegraphics[scale=0.9]{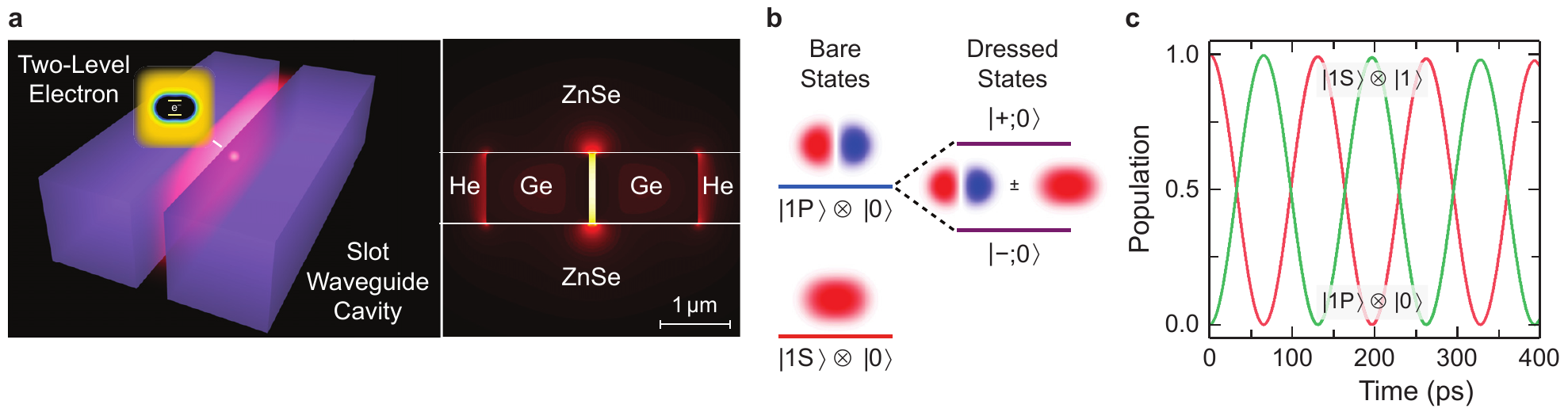}}
	\caption{Strong coupling and vacuum Rabi oscillations of a prolate electron bubble in a slot
		waveguide cavity. \textbf{a.} 3D structure of a slot waveguide cavity: Ge slot
		waveguide of 1~$\mu$m high with 100~nm slot clamped in ZnSe substrate, and
		calculated cross-sectional view of the electric field intensity in the cavity on-resonance with the 1S-to-1P transition wavelength. \textbf{b.} Formation of the true eigenstates, vacuum dressed states, by equal superpositions of the bare prolate 1S and 1P electron states.
		\textbf{c.} Calculated vacuum Rabi
		oscillations between the non-eigenstates 1S (photon-in-cavity) and 1P
		(photon-in-bubble) in a deformed electron bubble coupled with a single cavity
		photon. The initial state is the prolate 1P state with 0 photon in the
		cavity.}\label{Fig:VacuumRabiOscillation}
\end{figure*}

\noindent\textbf{Vacuum Rabi oscillations}

We now investigate strong coupling~\cite{scully1999quantum,boyd2003nonlinear,kimble1998strong,boca2004observation} between a single electron in the bubble and
zero or one photon in a cavity, on resonance at a mid-IR wavelength~\cite{soref2010mid,chen2014heterogeneously}. In view of
the necessary helium environment to hold the electron bubble, the most realistic
design for experiments is an on-chip slot waveguide cavity~\cite{hiscocks2009slot}, as illustrated in
Fig.~\ref{Fig:VacuumRabiOscillation}a. The slot serves as both a nanofluidic
channel and a photon-confinement channel~\cite{duh2012microfluidic}. The electron can be injected into He-4
by a field emission tip~\cite{golov1994ir,wei2016exotic} and be carried into the slot by the superfluid,
which feels zero resistance in arbitrarily small channels. The pressure can be
initially set at 25~bar, slightly below the liquid-solid phase transition, for
efficient electron loading. Afterwards, the pressure can be slightly increased
to transform the liquid into solid, which freezes the electron in place. An
ideal choice of materials for mid-IR slot waveguides are germanium (Ge) on zinc
selenide (ZnSe) (see Methods). Figure~\ref{Fig:VacuumRabiOscillation}a
shows the simulated field intensity distribution around two Ge rectangular
waveguides of dimensions 1~$\mu$m high and 1.5~$\mu$m wide with a 100~nm slot
filled with
helium. Photons are dominantly concentrated in the slot. To form a 3D cavity,
the waveguide length is cutoff at half of the propagation wavelength.

The eigenstates of a two-level electron coupled with a cavity of fixed number of
photons are the well-known dressed states~\cite{scully1999quantum,boyd2003nonlinear}. If the photon number is 0 when
the electron is in the excited state, and 1 when the electron is in the ground
state, then the eigenstates are vacuum dressed states, which read
\begin{align}
|+;0\rangle &= \frac{1}{\sqrt{2}} (|1\text{P}\rangle\otimes|0\rangle + |1\text{S}\rangle\otimes|1\rangle) ,\label{Eigenstate1}\\
|-;0\rangle &= \frac{1}{\sqrt{2}} (|1\text{P}\rangle\otimes|0\rangle - |1\text{S}\rangle\otimes|1\rangle),\label{Eigenstate2}
\end{align}
for on-resonance coupling. Fig.~\ref{Fig:VacuumRabiOscillation}b shows the relation between these states. The Rabi-split eigenenergies are $
E_{\pm} = \frac{1}{2}\hbar(\omega_{\text{1P}}-\omega_{\text{1S}}) \pm \frac{1}{2}\hbar\Omega
$, in which $
\Omega = \sqrt{(\omega_{\text{1P}}-\omega_{\text{1S}})^2 + 4g^2}
$
is the generalized vacuum Rabi frequency~\cite{scully1999quantum,boyd2003nonlinear},  $g=d\cdot\mathcal{E}(\bm{r}_0)/\hbar$ is the coupling frequency appearing in the
Jaynes-Cummings Hamiltonian, $d$ is the transition dipole moment, and
$\mathcal{E}(\bm{r}_0)$ is the electric field
strength of a single-photon wavefunction at the location $\bm{r}_0$~\cite{hiscocks2009slot}. $|0\rangle$ and $|1\rangle$ in Eqs.~(\ref{Eigenstate1}) and
(\ref{Eigenstate2}) denote the 0- and 1-photon states in the cavity. The
electron part suggests that the electron equally popularizes the 1S and 1P states,
the same as the equal superposition state in Fig.~\ref{Fig:LaserDrivenRabiOscillation}c. The bubble shape of these dressed states is thus the longlived prolate spheroid.

The prolate 1S and 1P electron states (bare states) have the eigenenergies
$0.249$~eV and $0.374$~eV from our DFT calclation. The 1S-to-1P transition energy is
0.125~eV, corresponding to a transition wavelength $ \lambda = 9.9$~um. The calculated transition dipole
moment between the prolate 1S and 1P states is 0.544~e~$\cdot$~nm. Our
electrodynamic calculation for our slot waveguide design gives a coupling
strength $g=2\pi\times 3.81$~GHz at the field maximum. Comparatively, the
spontaneous radiative decay of this two-level electron is
$\gamma_{\text{r}}=0.22$~MHz, \ie, $\tau_{\text{r}} = 4.5$~$\mu$s. The intrinsic
nonradiative decay in the solid He-4 matrix is $\gamma_{\text{nr}} \approx
0.1$~GHz, \ie,  $\tau_{\text{nr}}\approx 10$~ns ($10^3$ slower than in the
liquid~\cite{shikin1977mobility}). The cavity photon loss rate is $\kappa\approx 0.02$~GHz, \ie, a
quality factor $Q\approx 4.4\times10^6$.

Since in this system $g\gg\gamma_{\text{nr}}>\kappa>\gamma_{\text{r}}$ , the
system is in the electron-photon strong coupling regime. The cooperativity
$C=2g^2/(\kappa\gamma_{\text{nr}})\sim6\times10^5$ is extremely high~\cite{greuter2015towards,dovzhenko2018light}. If
we initialize the system on the non-eigenstates, either the ``photon-in-bubble"
state $|1\text{P}\rangle\otimes|0\rangle$ or the ``photon-in-cavity" state
$|1\text{S}\rangle\otimes|1\rangle$, we can observe longstanding population
oscillations, called the vacuum Rabi oscillations (or Rabi flopping), as a
result of the beating between two dressed eigenstates. In
Fig.~\ref{Fig:VacuumRabiOscillation}c, we can indeed see many cycles of vacuum
Rabi oscillations with almost indiscernible damping in hundreds of picoseconds.

\noindent\textbf{Two-electron entanglement}

When two electrons are put at the locations $\bm{r}_a$ and $\bm{r}_b$ in the same cavity, a single
photon can be shared by both electrons. The states of interest can be
written as the symmetric and antisymmetric entangled states, $|\text{SE}\rangle$
and $|\text{AE}\rangle$,  between the two
electrons,
\begin{align}
|\text{SE}\rangle &= \frac{1}{\sqrt{2}} \left(|1\text{S}\rangle_{\alpha} |1\text{P}\rangle_{\beta} \otimes|0\rangle + |1\text{P}\rangle_{\alpha} |1\text{S}\rangle_{\beta} \otimes|0\rangle \right),\\
|\text{AE}\rangle &= \frac{1}{\sqrt{2}} \left(|1\text{S}\rangle_{\alpha} |1\text{P}\rangle_{\beta} \otimes|0\rangle - |1\text{P}\rangle_{\alpha} |1\text{S}\rangle_{\beta} \otimes|0\rangle \right).
\end{align}
The $|1\text{S}\rangle_{\alpha} |1\text{P}\rangle_{\beta} \otimes |0\rangle$ and $|1\text{P}\rangle_{\alpha} |1\text{S}\rangle_{\beta} \otimes |0\rangle$ states represent
one photon absorbed in one electron bubble at $\alpha$ or $\beta$, and no photon left in
the cavity. They can be written as an equal superposition of $|\text{SE}\rangle$
and $|\text{AE}\rangle$. It has been shown~\cite{otten-prb-2015} that the
$|\text{SE}\rangle$ state
is a bright state, which is strongly coupled to the cavity, and the
$|\text{AE}\rangle$ state is a dark state, which is nearly uncoupled to the
cavity. We now demonstrate how a lossy cavity helps to efficiently build up the
entanglement between two electrons in a dark
state, similar to other dissipation-driven entangled
systems~\cite{Martin-Cano:11,otten-prb-2015,otten-pra-2016}. Such dissipative
production and stabilization protocols have been demonstrated in trapped
ion~\cite{lin2013dissipative} and superconducting
qubits~\cite{shankar2013autonomously}.

\begin{figure}[htb]
\centerline{\includegraphics[scale=0.9]{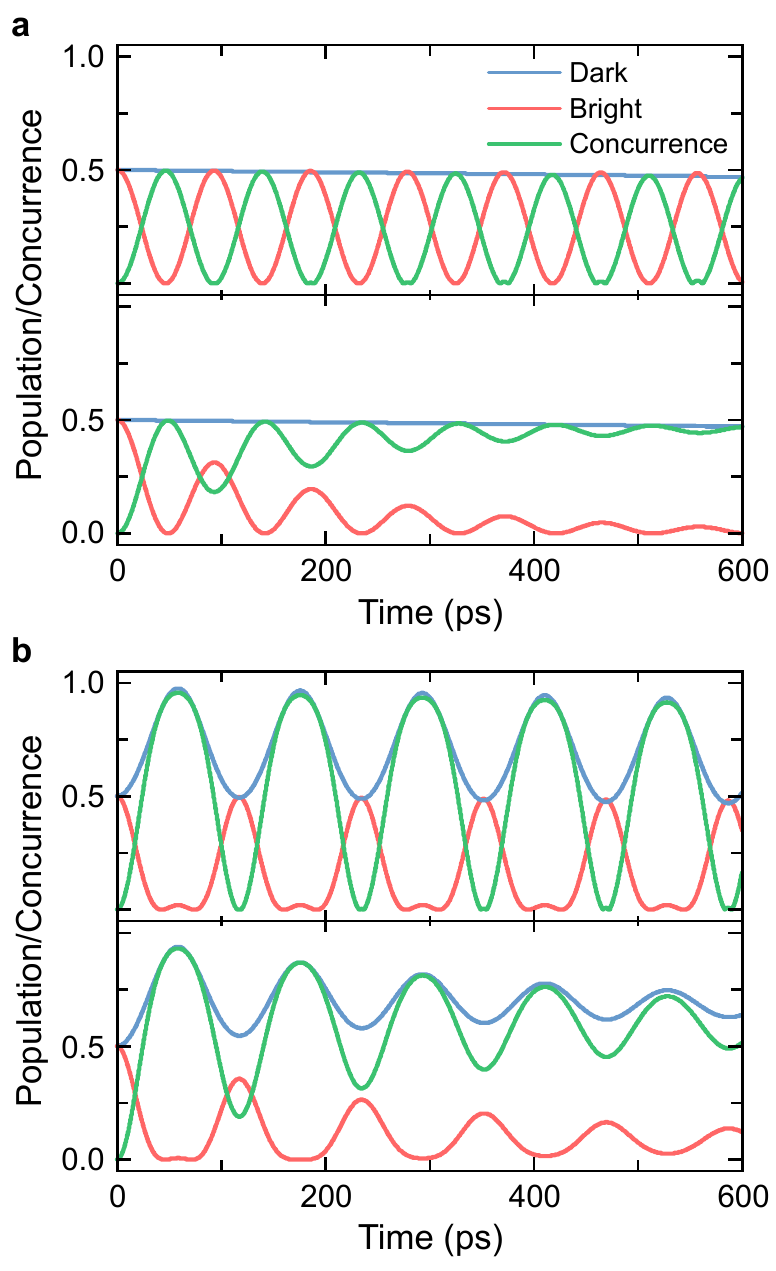}}
\caption{Evolution of entanglement between two prolate electron bubbles strongly
  coupled with a single photon in a slot-waveguide cavity. The photon frequency
  is on resonance with the 1S-to-1P transition. The
  initial state is a one-electron 1P state. Population of the (bright)
  symmetric entangled state, the (dark) antisymmetric entangled state, and
  the concurrence are plotted as a function of time. \textbf{a.} Evenly coupled
  electrons with $\kappa=0.02$~GHz
  (upper panel) or 10~GHz (lower panel) cavity photon loss rate. \textbf{b.}
  Unevenly coupled electrons with $\kappa=0.02$~GHz (upper panel) or 10~GHz
  (lower panel) cavity photon loss rate. }\label{Fig:TwoElectronEntanglement}
\end{figure}

Figure~\ref{Fig:TwoElectronEntanglement}a shows the
quantum dynamics of entangled states when two electron bubbles are equally
coupled, with $g=2\pi \times 3.81$~GHz, to a single cavity photon, starting from
the $|1\text{P}\rangle_{\alpha}  |1\text{S} \rangle _{\beta}\otimes |0\rangle$ state. The
intrinsic nonradiative decay is taken to be $\gamma_{\text{nr}}=0.1$~GHz, the
same as above. The
upper panel adopts a high-Q cavity with $\kappa=0.02$~GHz. Since the dark state
is a natural eigenstate of the system, uncoupled from the cavity, it maintains a
nearly constant population of 0.5. The bright state, on the other hand,
oscillates between 0.5 and 0, coupling into the cavity mode. The initial state
is completely unentangled, having a concurrence~\cite{Wootters:98} of 0. As the
bright state oscillates into the cavity, the concurrence grows, reaching a
maximum when the bright state reaches a population of 0. The lower panel adopts
a low-Q (highly-lossy) cavity $\kappa=10$~GHz. In this case, the bright state
oscillations are quickly damped out, leaving a long-lived entangled state from
the remaining dark state.

If we allow the two electron bubbles to have different couplings to the cavity,
typically by locating them at different points in the cavity, the amount of
entanglement can be enhanced. Fig.~\ref{Fig:TwoElectronEntanglement}b shows the
dynamics when the initially excited electron bubble has half the coupling
strength as the
other electron bubble. The dark state is now weakly coupled into the
cavity mode. This causes an indirect coupling between the bright and
dark states through the cavity, eventually leading to a large
gain in the antisymmetric population. The upper panel of
Fig.~\ref{Fig:TwoElectronEntanglement}b shows that the concurrence can reach
nearly 1, meaning maximally entangled, when the photon loss rate is low.
Increasing the photon loss in the cavity, as shown in the lower panel,
causes the oscillations to damp out quickly, but the final concurrence goes to
0.6, higher than the case when the two bubbles are equally coupled.
This photon-dissipation driven entanglement is not limited to two bubbles; if more were
coupled into the cavity, they would also become entangled in a genuine
multipartite way, where every electron bubble shares some degree of bipartite
entanglement with every other electron bubble~\cite{otten-pra-2016}. This type of
entanglement is similar to the $W$ state of quantum
information~\cite{dur-pra-2000}. The entangled states could form the basis of
interesting quantum information
technologies, such as highly-sensitive quantum sensors or generators of cluster
states for measurement-based quantum
computing~\cite{briegel-prl-2001,bell-njp-2013}.
Proposals for measuring such entanglement include looking at the bunching and
anti-bunching statistics of the system's emitted
photons~\cite{otten2019optical,dumitrescu2017antibunching}.

\noindent\textbf{Conclusion}

We have proposed a unique system of single electrons in condensed helium-4 as a new platform of quantum nodes in a quantum network. A single electron in helium can be coherently manipulated by mid-IR photons in both the strong-pulse regime and single-photon regime, and exhibit driven and vacuum Rabi oscillations. Two electrons can form entangled states through natural coupling with an optical cavity and show nanoseconds lifetime. Combined with the single-electron state control via laser pulses and the long lifetimes of the
entangled states, the electron-in-helium system offers an attractive avenue for quantum information science and technology.

\ \vspace{-0.5cm}\\

\noindent\textbf{Methods}

\small

\noindent\textbf{Density functional calculation}

We calculate the electronic structure in condensed He-4 using a
density-functional formalism~\cite{jin2010electrons,jin2010vortex}, which is a generalization to the Gross-Pitaevskii equation formalism in the Bose-Einstein condensation (BEC) systems~\cite{berloff2000capture}. Our density-functional formalism can accurately reproduce most properties of He-4 at near zero temperature, including: a. the equation of states and surface tension~\cite{dalfovo1995structural,stringari1987surface}; b. the 1~eV potential barrier of helium to electron at zero pressure~\cite{grau2006electron}; c. real-time vortex nucleation and sound emission~\cite{jin2010vortex,mateo2011excited}.

The energy functional of the combined electron-in-helium system is written as
\begin{equation}
\begin{split}
E[\psi,\varphi] = & \int \Dd\bm{r}\ \frac{\hbar^2}{2m_{\text{He}}} |\nabla\psi|^2 + \frac{\hbar^2}{2m_{\text{e}}} |\nabla\varphi|^2 \\
& + \frac{1}{2} g_2 |\psi|^4 + \frac{1}{3} g_3 |\psi|^6 + \frac{1}{4} g_4 |\psi|^8 \\
& + \frac{1}{2} w |\nabla|\psi|^2|^2 + f_0 |\varphi|^2|\psi|^2,
\end{split}
\end{equation}
where $\psi$ is the macroscopic wavefunction of helium normalized to the number
$N$ of helium atoms, and $\varphi$ is the wavefunction of a single electron. This
functional encloses the two-particle, three-particle, and four-particle
correlations between helium atoms through the $g_2, g_3, g_4$ terms and the surface energy through the $h_2$ term~\cite{dalfovo1995structural,stringari1987surface}. It approximates the electron-helium interaction at the s-wave scattering level by the $f_0=2\pi\hbar^2 l/m_{\text{e}}$ term~\cite{grau2006electron}.

The equations of motion consistent with this functional can be derived as
\begin{align}
\Ii\hbar\partial_t \psi =& -\frac{\hbar^2}{2m_{\text{He}}} \nabla^2 \psi + (g_2 |\psi|^2 + g_3 |\psi|^6 + g_4 |\psi|^8) \psi \\
& - h_2 \nabla(|\psi|^2) \psi + f_0 |\varphi|^2 \psi - \mu \psi \nonumber, \\
\Ii\hbar\partial_t \varphi =& -\frac{\hbar^2}{2m_{\text{e}}} \nabla^2 \varphi + f_0 |\psi|^2 \varphi,
\end{align}
where $\mu$ is the chemical potential which varies with pressure. These two coupled equations are highly nonlinear. Numerically evolving these equations in real or imaginary time (with a paralleled 4th-order finite-difference and 4th-order Runge-Kutta scheme) can exhibit remarkable quantum nonequilibrium and equilibrium behaviors~\cite{jin2010electrons,jin2010vortex}.

\noindent\textbf{Master equation calculation}

We calculate the quantum optical dynamics using the Lindblad master equation,
\begin{equation}\label{lindblad_me}
  \frac{d\rho}{dt} = -i [H, \rho] + L(\rho),
\end{equation}
where $\rho$ is the density matrix of the system, $H$ is the system Hamiltonian,
possibly including time-dependent terms,
and $L(\rho)$ are the Lindblad superoperators describing dissipation and decay.

\begin{figure}[htb]
	\centerline{\includegraphics[scale=0.8]{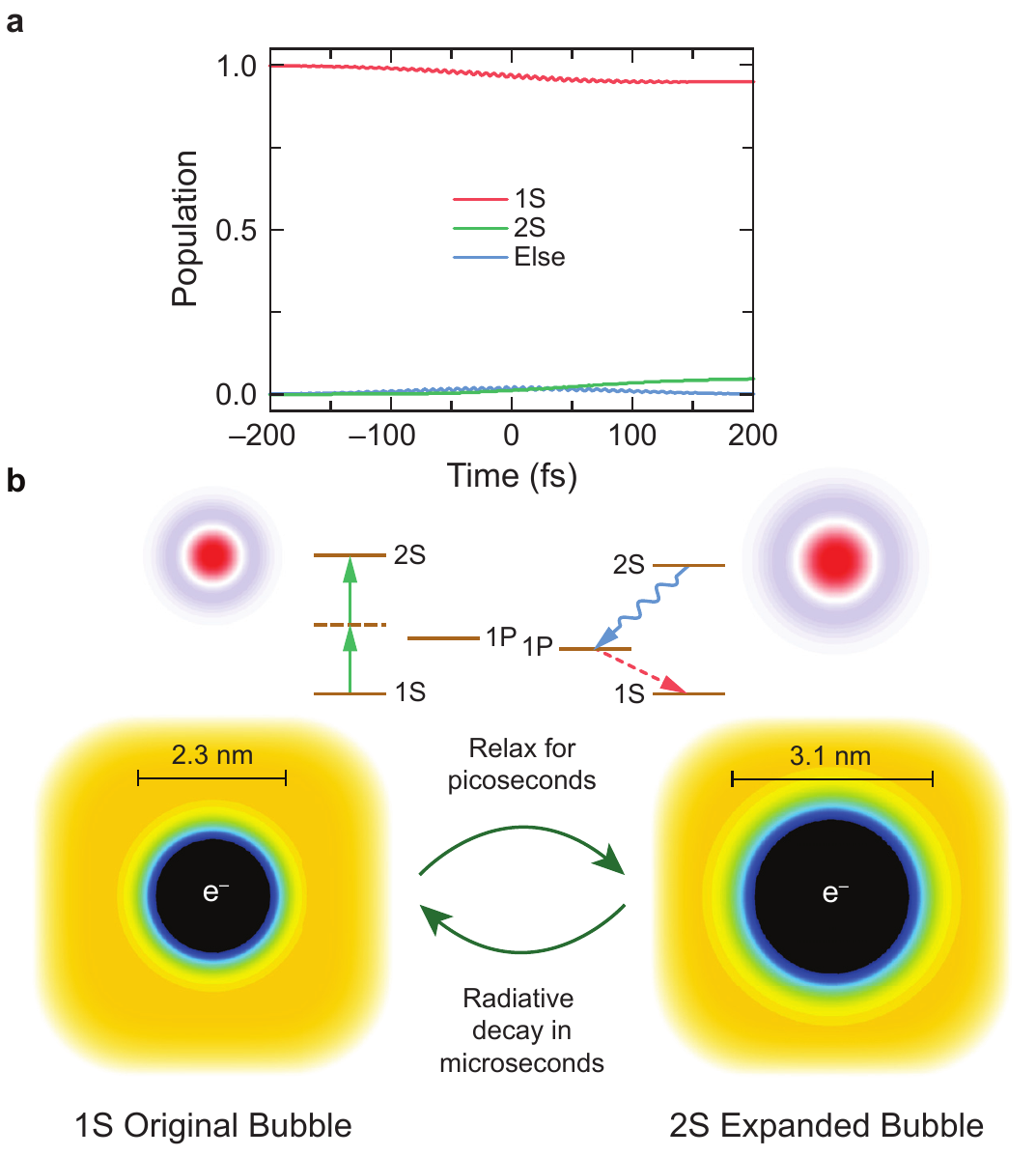}}
	\caption{Two-photon transition from the 1S to 2S state for a spherical bubble
		under 25~bar pressure and the fluorescence and bubble dynamics after the laser
		pulse. (a) Population of the 1S and 2S states for the pulse width
		$\mathcal{W}=200$~fs and electric field strength
		$\mathcal{E}=0.05$~V~nm$^{-1}$. (c) Processes of isotropic expansion of the
    2S bubble state after the excitation, the 2S-to-1P radiative decay and then
    nonradiative relaxation back to 1S bubble.}\label{Fig:TwoPhotonFluorescence}
\end{figure}

For the laser driven Rabi oscillations of
Fig.~\ref{Fig:LaserDrivenRabiOscillation}a-c, the Hamiltonian, $H$, is
defined by diagonal matrix elements, $H_0$, which represent the energy of each state
($|\text{1S}\rangle \langle \text{1S}|$ type terms) and off-diagonal matrix elements, $H_1$,
which represent the dipole moment between two states ($|\text{1S}\rangle \langle \text{1P}| +
|\text{1P}\rangle \langle \text{1S}|$ type terms). 48 states were included in the
Hamiltonian, approximating the infinite continuum. The values of the matrix
elements were computed using the density functional approach described above.
The total Hamiltonian was $H = H_0 + E(t) H_1$, where $E(t) =  G(t) \mathcal{E}
\cos(\omega_0 t)$, $G(t)$ is a Gaussian envelope such that the full width at
half maximum of $E^2(t)$ is the pulse length, $\mathcal{W}$, and the pulse
strength is $\mathcal{E}$.  The pulse
frequency, $\omega_0$, was on resonance with the $|\text{1S}\rangle$ to $|\text{1P}\rangle$
transition. Various pulse strengths ($\mathcal{E}$) and pulse lengths
($\mathcal{W}$) were used. Because of the strong laser driving in part of our
studies, our calculation generally does not presume the rotating wave
approximation (RWA). Hence our Rabi oscillations results may display
high-frequency oscillations on top. Due to the long radiative and non-radiative
lifetimes, no Lindblad terms were included in the simulations for
Fig.~\ref{Fig:LaserDrivenRabiOscillation}a-c. The simulation for
Fig.~\ref{Fig:TwoPhotonFluorescence}a used the same Hamiltonian but the frequency,
$\omega_0$, was on resonance with half the $|\text{1S}\rangle$ to $|\text{2S}\rangle$
transition frequency.

For the vacuum driven Rabi oscillations of Fig.~\ref{Fig:VacuumRabiOscillation}c,
the Hamiltonian describing the states of the electron bubble was truncated to
include just the $|1S\rangle$ and $|1P\rangle$ states and a cavity, also
truncated to two states, was added. The resulting Hamiltonian was, then,
\begin{equation}
  \begin{split}
  H = &\ \omega_{\text{1S}\rightarrow \text{1P}} |\text{1P}\rangle\langle \text{1P}| + \omega_{\text{1S} \rightarrow \text{1P}} a^\dagger a \\
  & + g (|\text{1S} \rangle\langle \text{1P}| + |\text{1P} \rangle\langle \text{1S}|) (a + a^\dagger),
 \end{split}
\end{equation}
where $\omega_{\text{1S}\rightarrow \text{1P}}$ is the $|\text{1S}\rangle$ to $|\text{1P}\rangle$
transition frequency, $a$ is the annihilation operator for the cavity, and $g$
is the electron bubble-cavity coupling of the slot waveguide design described in
Fig.~\ref{Fig:VacuumRabiOscillation}a. Spontaneous emission of the two-level
electron system was included via a Lindblad term, $L_{e^-}(\rho) = \gamma_\text{r}
\mathcal{D}[|\text{1P}\rangle\langle \text{1S}|](\rho)$, where $\mathcal{D}[C]\rho = C \rho
C^\dagger - \frac{1}{2} (C^\dagger C \rho + \rho C^\dagger C)$. Cavity photon
loss was included via Lindblad term
$L_c(\rho) = \kappa \mathcal{D}[a](\rho)$. Non-radiative decay, which is on the
order of
nanoseconds, was not included due to its long time scale. The system was
initialized in the $|\text{1S}\rangle \otimes |1\rangle$ state, representing one photon
in the cavity.

For the entanglement generation of Fig.~\ref{Fig:TwoElectronEntanglement}, an
additional two-level electron bubble was added to the system. The resulting
Hamiltonian was
\begin{equation}
  \begin{split}
  H =\ & \omega_{\text{1S}\rightarrow \text{1P}} |\text{1P}\rangle_{\alpha} {}_{\alpha}\langle \text{1P}| + \omega_{\text{1S}\rightarrow \text{1P}} |\text{1P}\rangle_{\beta} {}_{\beta}\langle \text{1P}| \\
  & + \omega_{\text{1S}\rightarrow \text{1P}} a^\dagger a \\
  & + g_{\alpha} (|\text{1S}\rangle_{\alpha} {}_{\alpha}\langle \text{1P}| + |\text{1P} \rangle_{\alpha} {}_{\alpha}\langle \text{1S}|) (a + a^\dagger)  \\
  & + g_{\beta} (|\text{1S}\rangle_{\beta} {}_{\beta}\langle \text{1P}| + |\text{1P} \rangle_{\beta}{}_{\beta}\langle \text{1S}|) (a + a^\dagger),
  \end{split}
\end{equation}
where the subscripts ${\alpha}$ and ${\beta}$ represent the different electron bubbles. In
this Hamiltonian, each bubble is coupled to the cavity (potentially with
different strengths) but the bubbles are not coupled to each other. Spontaneous
emission was included for each electron bubble using $L_{{\alpha},\text{r}} = \gamma_\text{r}
\mathcal{D}[|\text{1P}\rangle_{\alpha} {}_{\alpha}\langle \text{1S} |](\rho)$ and $L_{{\beta},\text{r}} = \gamma_\text{r}
\mathcal{D}[|\text{1P}\rangle_{\beta}{}_{\beta}\langle \text{1S} |](\rho)$. Non-radiative emission was
included via $L_{{\alpha},\text{nr}} =
\gamma_{\text{nr}}
\mathcal{D}[|\text{1P}\rangle_{\alpha} {}_{\alpha}\langle \text{1S} |](\rho)$ and $L_{{\beta},\text{nr}} = \gamma_{\text{nr}}
\mathcal{D}[|\text{1P}\rangle_{\beta} {}_{\beta}\langle \text{1S} |](\rho)$. Cavity loss was also included
(see above).

\noindent\textbf{Material choice for mid-IR slot waveguides}

Ge and ZnSe both are widely transparent in the 1 to 12~$\mu$m
wavelength range~\cite{kim2018verification,macdonald2013efficient} and have almost perfectly matched lattice constant (5.658~{\AA}
versus 5.668~{\AA})~\cite{calow1968growth}, which is superior for molecule beam epitaxy (MBE) growth
and wafer bonding process. Furthermore, ZnSe is transparent in the visible
wavelength down to 600~nm wavelength, particularly convenient for optical
alignment. Compared with vacuum
($\epsilon_{\text{vac}}=1$) and helium ($\epsilon_{\text{He}}=1.057$), both
materials have a large dielectric contrast ($\epsilon_{\text{Ge}}=15.5$
and $\epsilon_{\text{ZnSe}}=5.76$), especially good for on-chip light
confinement. To experimentally realize the slot waveguide structure shown in the
main text, one can use photolithography and dry etching to first produce a
3~$\mu$m wide ridge on the substrate, and then use a focused ion beam (FIB) to
make a 100~nm wide
1~$\mu$m deep slot in the middle. Photons can be sent into the cavity by direct illumination on top of the chip.

\noindent\textbf{Scheme of two-photon fluorescence imaging}

In practice, to precisely locate and optically manipulate a single electron in condensed helium, we need to develop a confocal
microscopy based on a certain fluorescence channel, similar to the technique used for defect centers in diamond~\cite{gruber1997scanning}.

The 1-photon 1S-to-1P transition shown in Fig.~3 is not an ideal fluorescence channel either in liquid or solid helium. The complex shape deformation and phonon emission sensitively depend on the laser pulse width and strength, and can cause completely quenched or overly broadened fluorescence spectrum. Instead, the 2-photon 1S-to-2S transition is a better channel for imaging. As
sketched in the inset of Fig.~\ref{Fig:BubbleEnergyAndTransition}, the electron first jumps into a
virtual level between the 1S and 2S state~\cite{boyd2003nonlinear} and then into the 2S state, by sequentially absorbing two photons at the
frequency $\omega = (\omega_{2\text{S}}-\omega_{1\text{S}})/2$. Afterwards, the
electron makes a 1-photon 2S-to-1P transition by the spontaneous emission and
produces a fluorescence photon at a fixed wavelength.

For most systems under conventional laser illumination,
the 2-photon transition rate is low	~\cite{boyd2003nonlinear}. However, we find that the pulse energy from
a parametrically amplified femtosecond laser provides a decent probability to
bring the electron into the 2S state. The 2-photon transition wavelength at
$p=25$~bar pressure from our calculation is $\lambda=4.08$~$\mu$m.
As shown in Fig.~\ref{Fig:TwoPhotonFluorescence}a, after a single laser pulse of
$W=200$~fs and $\mathcal{E}=0.05$~V~nm$^{-1}$, the electron has about 5\%
population on the 2S state. There is little leakage to other states (mainly the nearby 1P state). This transition rate is sufficient to give strong signal on a mid-IR photodetector~\cite{norton2002hgcdte}.

Interestingly, before the 2S-to-1P radiative decay sets in, the 2S bubble has to undergo a nonradiative isotropic expansion first, as illustrated in Fig.~\ref{Fig:TwoPhotonFluorescence}c~\footnote{Several
	authors adopted a sharp-wall infinite-barrier model to calculate the stability
	of 2S bubble and found that it evolves into a tetrahedra shape~\cite{maris2007calculation,grinfeld2003instability}. However,
	we have performed a more accurate 3D DFT calculation, and found no indication
	of such symmetry breaking.}. This is due to the increased quantum pressure of
2S electron wavefunction exerted on the bubble surface. Our  calculation shows that the
bubble radius increases from 1.14~nm to 1.54~nm in a time scale about 10~ps. The
final 2S state is metastable and does not exhibit further deformation or nonraditative decay. It decays only through the spontaneous emission into the 1P state in the expanded bubble. The 1-photon fluorescence wavelength for the expanded 2S to 1P state is $\lambda=5.13$~$\mu$m, and the
dipole moment is $d=0.31$~e$\cdot$nm. The spontaneous emission rate can be calculated to be $\gamma=5.20\times10^5$~s$^{-1}$, \ie, lifetime $\tau = 1.92$~$\mu$s. After
the 2S-to-1P transition, the bubble will likely undergo a similar fission
process as Fig.~\ref{Fig:LaserDrivenRabiOscillation}a, and finally relax back to
the original 1S ground state. This final nonradiative relaxation from the 1P
state back to 1S state has no impact on the already emitted photon. Hence the
2-photon 1S-to-2S transition provides a reliable approach for confocal imaging
single electrons in condensed helium.

\  \\
\noindent\textbf{Data availability} 

All relevant data are available from the authors on request.

\noindent\textbf{Code availability} 

All relevant codes are available from the authors on request.

\  \\
\noindent\textbf{Acknowledgments}

This work was performed at the Center for Nanoscale Materials, a U.S. Department of Energy Office of Science User Facility, and supported by the U.S. Department of Energy, Office of Science, under Contract No. DE-AC02-06CH11357.

\bibliography{ElectronBubbleRef}

\end{document}